\documentclass[aps,pra,superscriptaddress,amsmath,amssymb,preprintnumbers,floatfix,showpacs,11pt]{revtex4}

\usepackage{amssymb}
\usepackage{epsfig}
\usepackage{subfigure}
\usepackage{graphicx}

\begin{document}
\title{Robustness of nonclassical superpositions states against decoherence}
\author{Faisal A. A. El-Orany}
\affiliation{ Department of Mathematics and Computer Science,
Faculty of Natural Science, Suez Canal University, Ismailia,
Egypt; Department of Optics, Palack\'y University,
17.~listopadu~50, 772 07~Olomouc, Czech Republic }

\date{\today}

\begin{abstract}
We make a comparative study of quadrature squeezing, photon-number
distribution and Wigner function in a decayed quantum system.
Specifically, for a field mode prepared initially in cat states
interacting with a zero-temperature environment, we show that the
rate of reduction of the nonclassical effects in this system is
proportional to the occurrence of the decoherence process.
\end{abstract}

  \maketitle
\section{Introduction}

Decoherence process represents the transformation of superposition
states into statistical mixture states, i.e. the off-diagonal elements of
the system are suppressed. This can occur through, e.g., the interaction between
system and environment. Actually, the decoherence process is important not
only for understanding the quantum-classical transition \cite{zurek2}, but
also it may eventually be useful for applications that require keeping
coherence in mesoscopic or macroscopic systems, such as quantum computation
\cite{quan}. Furthermore, the decoherence is at the heart of the quantum theory
of measurement \cite{zurek1}.

On the other hand,  superposition principle is at the heart of the
quantum mechanics. It implies that probability densities of
observable quantities usually exhibit interference effects instead
of simply being added. The most significant examples reflecting
the power of such principle are the Schr\"{o}dinger cat states
\cite{sch1}, which exhibit various nonclassical effects, such as
squeezing, sub-Poissonian statistics and oscillations in
photon-number distribution \cite{yur,yur1,micro1}, even if their
components are close to the classical ones \cite{buz1},i.e., they
are minimum uncertainty states and exhibit Poissonian
distribution. These states can be defined as:

\begin{equation}
|\alpha \rangle _{\phi }=A^{\frac{1}{2}}[|\alpha \rangle +\exp (i\phi
)|-\alpha \rangle ],  \label{1}
\end{equation}
where $|\alpha \rangle $ is a coherent state with complex amplitude $\alpha $
, $\phi $ is a relative phase and $A$ is the normalization constant having
the form

\begin{equation}
A=\frac{1}{2[1+ \exp(-2|\alpha|^{2})\cos \phi]}.  \label{2}
\end{equation}
Specifically,  for $\phi=0,\pi$ and $\pi/2$ state  (\ref{1}) reduces to even
coherent (ECS), odd coherent (OCS) and
Yurke-Stoler (YSS) states, respectively. It is worth mentioning that
 there are two regimes controlling the behavior of the states (
\ref{1}), which are microscopic regime
for small values of $|\alpha|$
(i.e., when the "distance" between the components of the cat is small)
and macroscopic regime for large values of $|\alpha|$ \cite{faisal2}.
In fact,
these states are more nonclassical in the microscopic regime.
In other words, the amount of nonclassical effects, such as the negative
values in Wigner function and the oscillatory behavior in the photon-number
distribution for these types of states are more pronounced in the microscopic
regime than in the macroscopic regime.
For more
details about states (\ref{1}), such as their generations and their
properties when they are evolving in various optical systems, one can consult
the review article \cite{bivi}, and references therein.

In this article we study the relation between the decoherence process and
the occurrence of the nonclassical effects in a decayed quantum system. More
precisely, we  compare development of
nonclassical effects in both  quadrature squeezing and photon-number
distribution with  the occurrence of interference pattern in the Wigner
function.
We perform  such a comparison for the field mode
prepared initially in the state (\ref{1}) (described by the
density matrix $\hat{\rho}$) which interacts with zero-temperature
environment. The master equation in the Born-Markov approximation
 describing the  system is \cite{milb}

\begin{equation}
\frac{\partial \hat{\rho}}{\partial t}=\frac{\gamma }{2}(2\hat{a}\hat{\rho}%
\hat{a}^{\dagger }-\hat{a}^{\dagger }\hat{a}\hat{\rho}-\hat{\rho}\hat{a}
^{\dagger }\hat{a}),  \label{2c}
\end{equation}
where $\gamma $ is the decay constant and $\hat{a}\quad (\hat{a}^{\dagger })$
is the annihilation (creation) operator designated to the mode of the field.
The well-known time dependent solution for (\ref{2c}) is \cite{barnett}

\begin{equation}
\hat{\rho}(t)=A\sum\limits_{j,j^{\prime}=1}^{2}\exp(i\phi_{jj^{\prime}})
\langle \alpha_{j}|\alpha_{j^{\prime}}\rangle^{1-\mu} |\sqrt{\mu}%
\alpha_{j}\rangle \langle\sqrt{\mu}\alpha_{j^{\prime}}|,  \label{2a}
\end{equation}
where $\alpha_{1}=\alpha,\quad \alpha_{2}=-\alpha$ , $\mu=\exp(-\tau),\quad
\tau=t\gamma$ is the scaled decaying parameter and
\begin{equation}
\phi_{jj^{\prime}}=\left\{
\begin{array}{rl}
0 \;\;\;  {\rm for} \;\; j=j^{\prime}, \\
\phi\;\;\;  {\rm for} \;\; j>j^{\prime}, \\
-\phi\;\;\;  {\rm for} \;\; j<j^{\prime}.
\end{array}
\right.  \label{2b}
\end{equation}

\section{Quadrature squeezing}

As  is well known  squeezing is one of the most important phenomena in quantum
optics because of its applications in various areas, e.g., in optics
communication, quantum information theory, etc. \cite{inf1}. Squeezed light
can be measured by a homodyne detection where the signal is superimposed on
a strong coherent beam of the local oscillator.

Here we investigate  quadrature squeezing for the density matrix (\ref{2a}
). For this purpose we define the position and momentum operators, which are
related to the conjugate electric and magnetic field operators $\hat{E}$ and
$\hat{H}$ of electromagnetic waves, as
\begin{equation}
\hat{X}=\frac{1}{2}(\hat{a}+\hat{a}^{\dagger }),\qquad \hat{Y}=\frac{1}{2i}(
\hat{a}-\hat{a}^{\dagger }),  \label{2d}
\end{equation}
where $[\hat{X},\hat{Y}]=\frac{i}{2}$; then the uncertainty relation reads $
\langle (\triangle \hat{X})^{2}\rangle \langle (\triangle \hat{Y}
)^{2}\rangle \geq \frac{1}{16}$ where $\langle (\triangle \hat{X}
)^{2}\rangle =\langle \hat{X}^{2}\rangle -\langle \hat{X}\rangle ^{2}$.
Therefore, we can say that the mode is squeezed if $S_{1}(t)=4\langle
(\triangle \hat{X})^{2}\rangle -1<0$ or $S_{2}(t)=4\langle (\triangle \hat{Y}
)^{2}\rangle -1<0$.

\begin{figure}
  \centering
  {\includegraphics[width=8cm]{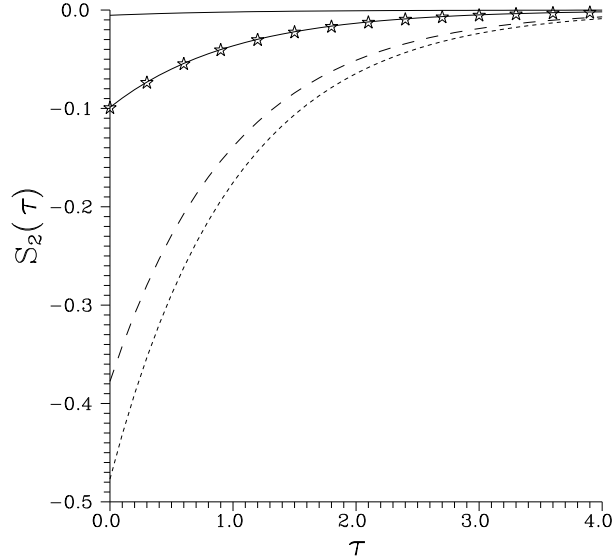}}
 \caption{ Squeezing factor $S_{2}(\tau)$ of ECS case against the
scaled decaying parameter $\tau$ for $\alpha=2$ (solid curve),
$1.5$ (star-centered curve), $1$ (short-dashed curve) and $0.5$
(long-dashed curve). }
\end{figure}

Now squeezing factors $S_{j}(t)$
for the system under
consideration---restricting ourselves to ECS case---take the forms

\begin{equation}
S_{1}(t)=\mu S_{1}(0)=\frac{4\mu \alpha ^{2}}{1+\exp (-2\alpha ^{2})},
\quad
S_{2}(t)=\mu S_{2}(0)=\frac{-4\mu \alpha ^{2}\exp (-2\alpha ^{2})}{1+\exp
(-2\alpha ^{2})},\label{3a}
\end{equation}
where $S_{j}(0)$ are the initial squeezing factors. We have considered
here  $\alpha $ to be real. From these expressions it is clear
that the quantum fluctuation of the field decreases exponentially as a
result of its interaction with environment. More precisely, $S_{j}(t)$ decay at
the same rate as the intensity of the field \cite{nig}. Further, we see that
squeezing exists provided that $\alpha $ and $\tau $ are finite. Of
course, the origin of these nonclassical effects is in the interference
between the components of the cat. One can also check that when $\tau $ is
large enough, squeezing factors $S_{j}(t),j=1,2$ tend to zero. In other words, the
system tends to a steady state, which, in this case, is a pure state (vacuum
state). In Fig. 1 we  plot $S_{2}(\tau )$  for shown values of the
parameters. Generally we can see that squeezing is more pronounced in the
microscopic regime (when $\alpha $ is small). Furthermore,
for $\alpha \geq 2$ the squeezing factor tends to zero regardless
of the values of $\tau $.
In this case the system
becomes a statistical mixture state or a vacuum state
if the values of $\tau$ are small or large.
 This point will be  clear
when investigating the behavior of the Wigner function in section IV.

\section{Photon-number distribution}

Photon-number distribution $P(n)$ is an integral part of the modern
description of light, which can be measured by photon detectors based on the
photoelectric effect. Further, one of the most interesting nonclassical
effects emerging from the superposition principle is the oscillatory
behavior in $P(n)$. In general, such behavior is closely related to that
of the Wigner function, however, this is a necessary but not sufficient
condition. For example, the $P(n)$ of ECS, OCS and YSS are completely
different; whereas those of ECS and OCS exhibit pairwise oscillations in
phase space (even number of photons can be observed for ECS and odd numbers
for OCS), the distribution of YSS is a Poissonian even though the behavior
of the Wigner function for these states is qualitatively similar. The second
issue we want to address here is that in general the occurrence of squeezing
in the quadrature variances does not need to be accompanied by oscillations
in the $P(n)$ and vice versa. For instance, binomial states \cite{kni} can
exhibit quadrature squeezing even though their $P(n)$ are close to
Poissonian ones. In the same spirit, for ECS the oscillations in $P(n)$ are
more pronounced when the ''distance'' between the basis of the cat increases,
however, this is not the case for the quadrature squeezing, which is completely
suppressed for $\alpha \geq 2$ (see Fig. 1).
\begin{figure}
\centering
  \subfigure[]{\includegraphics[width=8cm]{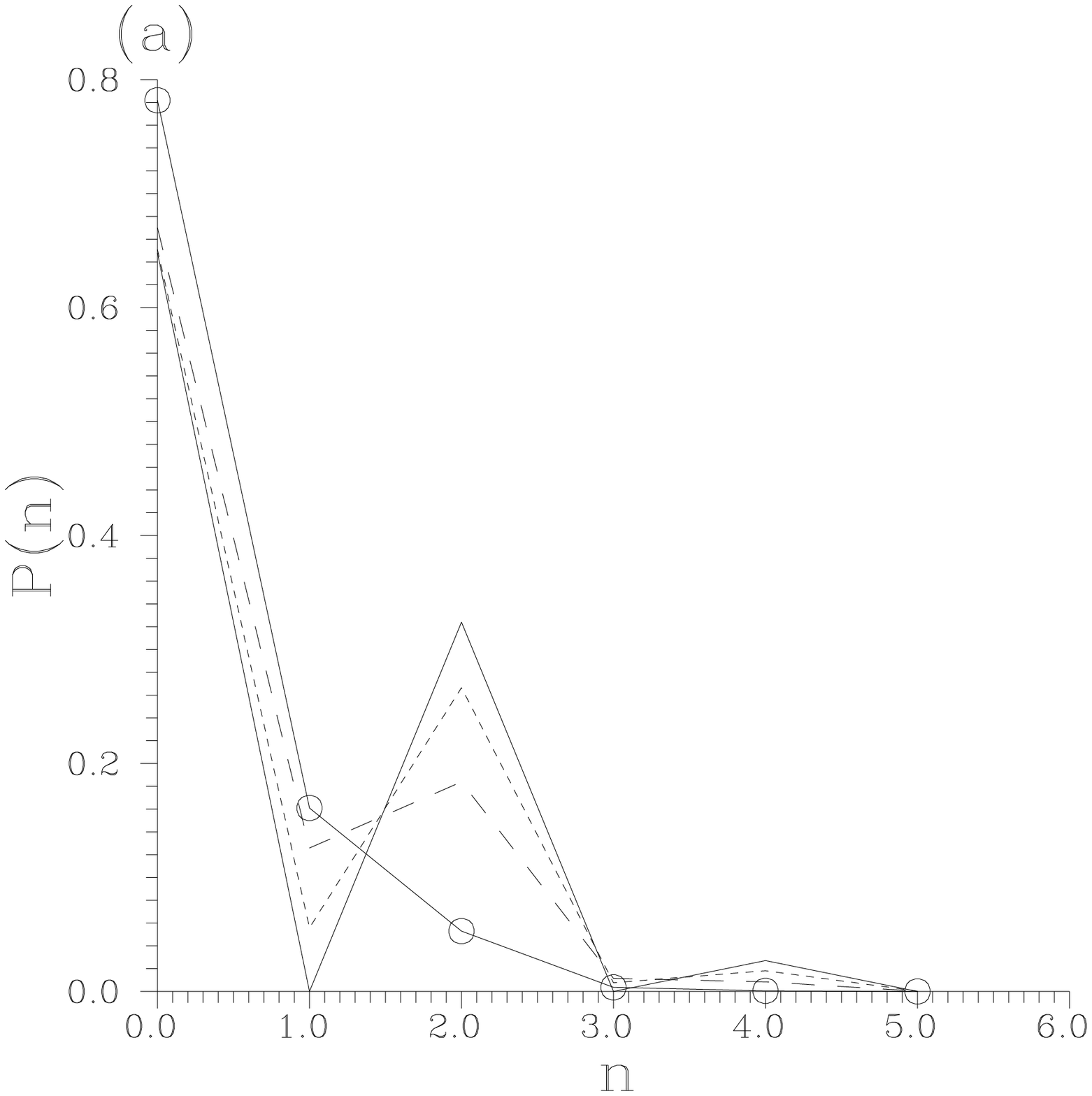}}
 \subfigure[]{\includegraphics[width=8cm]{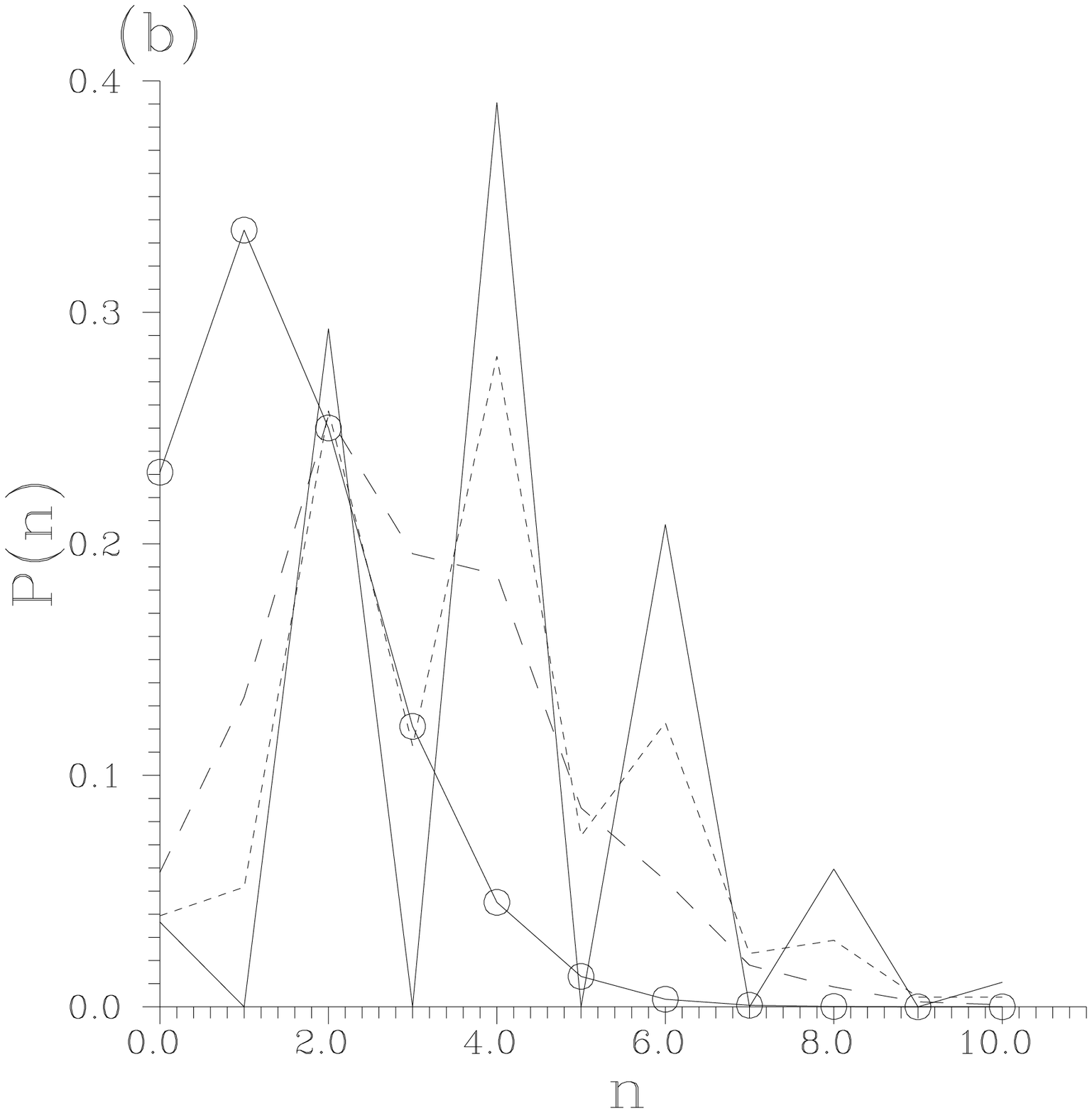}}
\caption{$P(n)$ of ECS case against $n$ for $\alpha=1$ (a), $2$
(b) and for $\tau=0$ (solid curve), $0.1$ (short-dashed curve),
$0.3$ (long-dashed curve) and $1$ (circle-centered curve). }
\end{figure}
Now we investigate the sensitivity of the $P(n)$ of the system
under consideration to lossy mechanism. This quantity can be
calculated easily ($P(n)=\langle n|\hat{\rho}(t)|n\rangle$) and
one obtains
\begin{equation}
P(n)=2A\frac{(\sqrt{\mu }\alpha )^{2n}}{n!}\exp (-\mu \alpha ^{2})\left\{
1+(-1)^{n}f(\alpha )\cos \phi \right\} ,  \label{4a}
\end{equation}
where
\begin{equation}
f(\alpha )=\exp [-2\alpha ^{2}(1-\mu )].  \label{4b}
\end{equation}
By comparing the expression (\ref{3a}) with (\ref{4a}) we find that the
dissipation is involved in two quantities by different ways and consequently
the sensitivity of these quantities to lossy mechanism is completely
different. As before, the origin of the nonclassical oscillations in the $%
P(n) $ lies in the interference in phase space. Further, in (\ref{4a}) the
interference term is decaying by the factor $f(\alpha )$ \cite{milb} and
thus its contribution is more pronounced---oscillatory behavior can occur in
$P(n)$---when $\alpha $ and $\mu $ are small. This situation is  similar to
that of the quadrature
squeezing.  We will discuss this point quantitatively in section IV by
investigating the
behavior of the factor $f(\alpha)$. In Figs. 2 we plot the $P(n)$ for ECS against $n$
for microscopic $(a)$ and macroscopic $(b)$ regimes, respectively, for given
values of the parameters. By comparing  the curves in Fig. 2a with those having
the same values of $\tau $ in Fig. 2b, one can conclude that the
oscillations in $P(n) $ for macroscopic regime are suppressed
faster than  those for microscopic regime provided that $\tau $ is small.
Also the comparison between the behavior of both the short-dashed curves in  Fig. 1
and in Fig. 2a shows that the $P(n)$ is more sensitive to dissipation than the
quadrature squeezing is. This is clear as one can observe that the oscillations
in the $P(n)$ are completely suppressed, however, squeezing is still
remarkable in the quadrature squeezing. The final remark is that the
behavior of the $P(n)$ when $\tau =0.3,1$ in Figs. 2 is close to
that for
the statistical mixture of coherent states under the influence of the decay
mechanism.

\section{Wigner function}

Wigner ($W$) function is one of the quasiprobability functions, which
carries  full information about the quantum system. This function is
sensitive to the interference in phase space and can be
realized in optical homodyne tomography \cite{tom}. Here we use this
function to study the decoherence of the system under discussion. The
definition of the decoherence has been given in the Introduction.

 The $W$ function can be defined as
\begin{equation}
W(\beta ,t)=\frac{1}{\pi ^{2}}\int d^{2}\zeta \exp (\beta \zeta ^{\ast
}-\beta ^{\ast }\zeta )C^{(w)}(\zeta ,t),  \label{21h}
\end{equation}
where $C^{(w)}(\zeta ,t)$ is the symmetrically ordered characteristic
function having the form

\begin{equation}
C^{(w)}(\zeta ,t)={\rm Tr}[\hat{\rho}(t)\exp (\hat{a}^{\dagger }\zeta -%
\hat{a}\zeta ^{\ast })],  \label{21ha}
\end{equation}
where $\hat{\rho}(t)$ is the density matrix of the system, which for the
system
under consideration is given by (\ref{2a}). For the future purpose, we
derive the $W$ function  following the same steps as in \cite{buz1}.
Thus we rewrite the $W$ function in terms of the normally ordered moments of
the creation and annihilation operators using the Baker-Hausdorff theorem.
Therefore (\ref{21h}) takes the form
\begin{equation}
W(\beta ,t)=\frac{1}{\pi ^{2}}\sum\limits_{n,m=0}^{\infty }
\frac{\langle\hat{a}^{\dagger m}(t)\hat{a}^{n}(t)\rangle }{n!m!}
I_{\rm mn},\label{21ab}
\end{equation}
where we have used the abbreviation
\begin{eqnarray}
\begin{array}{lr}
I_{\rm mn}=
\int d^{2}\zeta \exp (-\frac{1}{%
2}|\zeta |^{2}+\zeta ^{\ast }\beta -\zeta \beta ^{\ast })\zeta ^{m}(-\zeta
^{*})^{n}                                  \\
\\
\equiv (-1)^{n+m}\frac{\partial ^{m+n}}{
\partial \beta ^{\ast m}\partial \beta ^{n}}
\int d^{2}\zeta \exp (-\frac{1}{
2}|\zeta |^{2}+\zeta ^{\ast }\beta -\zeta \beta ^{\ast }).
\label{21ac}
\end{array}
\end{eqnarray}
Carrying  out the integration in
(\ref{21ac}) we obtain
\begin{equation}
I_{\rm mn}
=2\pi (-1)^{n+m}\frac{\partial ^{m+n}}{
\partial \beta ^{\ast m}\partial \beta ^{n}}
\exp (-2|\beta |^{2}). \label{21ad}
\end{equation}
After minor algebra and using  the
Rodrigues' formula for Laguerre polynomial, (\ref{21ad}) reads
\begin{equation}
I_{\rm mn}
=2^{n+1}\pi (-1)^{m} \beta ^{*(n-m)}m! {\rm L}^{n-m}_{m}(2|\beta| ^{2})
\exp (-2|\beta |^{2}), \label{21ae}
\end{equation}
where $L^{k}_{m}(.)$ are the associated Laguerre polynomials of order $m$.

On the other hand, the normally ordered expectation values
$\langle\hat{a}^{\dagger m}(t)\hat{a}^{n}(t)\rangle$ associated
with  the density matrix (\ref{2a}) are given   as \cite{phon}:
\begin{equation}
\langle\hat{a}^{\dagger m}(t)\hat{a}^{n}(t)\rangle=
A\sum\limits_{j,j^{\prime }=1}^{2}\exp (i\phi _{jj^{\prime }})\langle \alpha
_{j}|\alpha _{j^{\prime }}\rangle \alpha _{j}^{n}\alpha _{j^{\prime }}^{\ast
m} \mu^{\frac{m+n}{2}}.   \label{elor1}
\end{equation}

On substituting (\ref{21ae}) and (\ref{elor1}) into (\ref{21ab})
we arrive at

\begin{equation}
W(\beta ,t)
=\frac{A\exp (-2|\beta |^{2})}{\pi }\sum\limits_{n,m=0}^{\infty }\left[
\sum\limits_{j,j^{\prime }=1}^{2}\exp (i\phi _{jj^{\prime }})\langle \alpha
_{j}|\alpha _{j^{\prime }}\rangle \alpha _{j}^{n}\alpha _{j^{\prime }}^{\ast
m}\right] \frac{(-1)^{m}2^{n+1}\mu ^{\frac{n+m}{2}}}{n!\beta ^{\ast (m-n)}}%
{\rm L}_{m}^{n-m}(2|\beta |^{2}).
\label{21a}
\end{equation}
On using  the generating function for Laguerre polynomials
and  the Taylor's expansion for the exponential function, (\ref{21a}) reduces
to the following closed form

\begin{equation}
W(\beta ,t)
=\frac{2A}{\pi }\Bigl[\exp (-2|\beta -\alpha \sqrt{\mu }|^{2})+\exp
(-2|\beta +\alpha \sqrt{\mu }|^{2})
+2f(\alpha )\cos \phi \cos (4\alpha \sqrt{\mu }{\rm Im}\beta )\exp
(-2|\beta |^{2})\Bigr],
\label{21b}
\end{equation}
where $f(\alpha )$ is given by (\ref{4b}).

\begin{figure}
\centering
  \subfigure[]{\includegraphics[width=8cm]{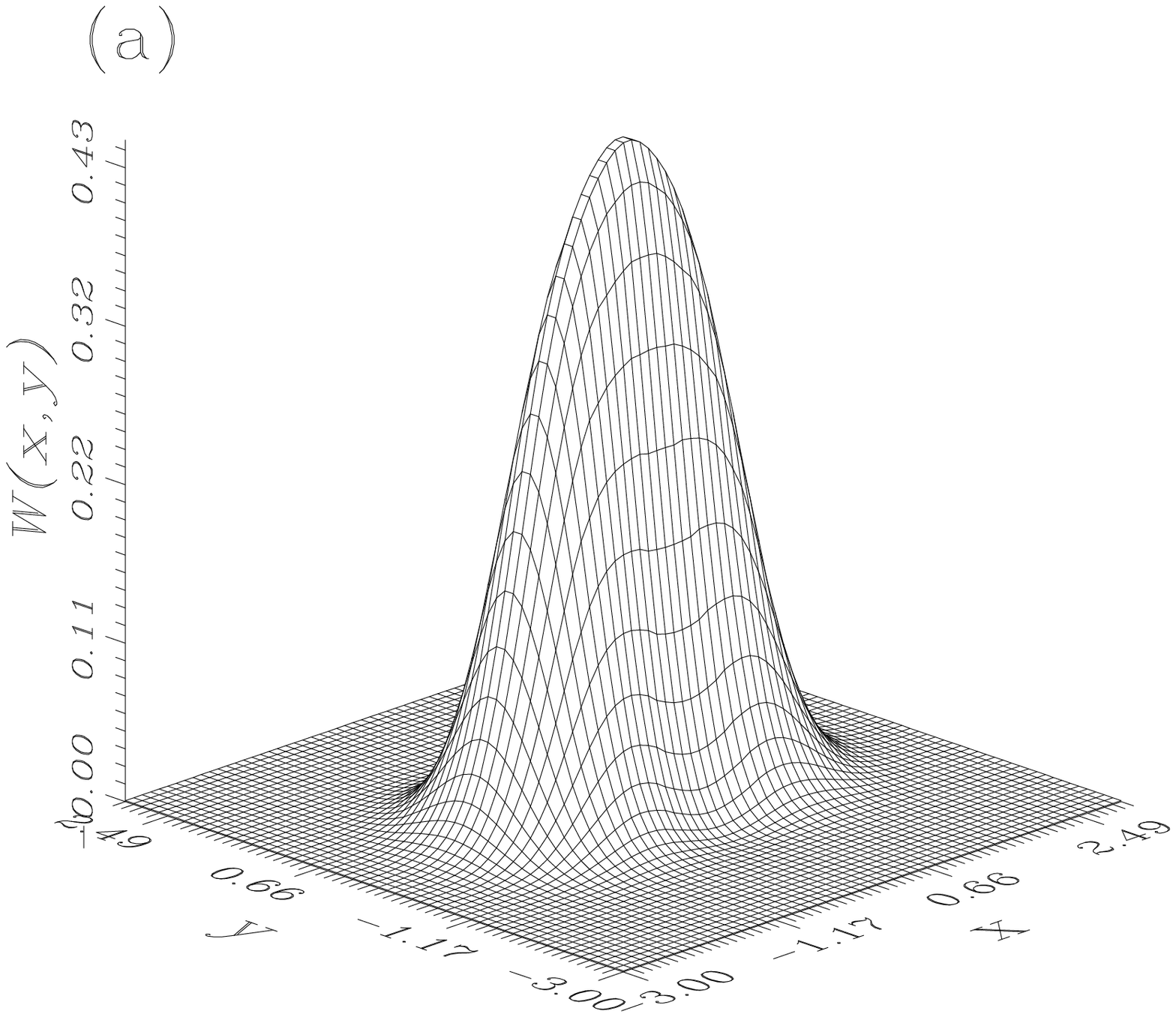}}
 \subfigure[]{\includegraphics[width=8cm]{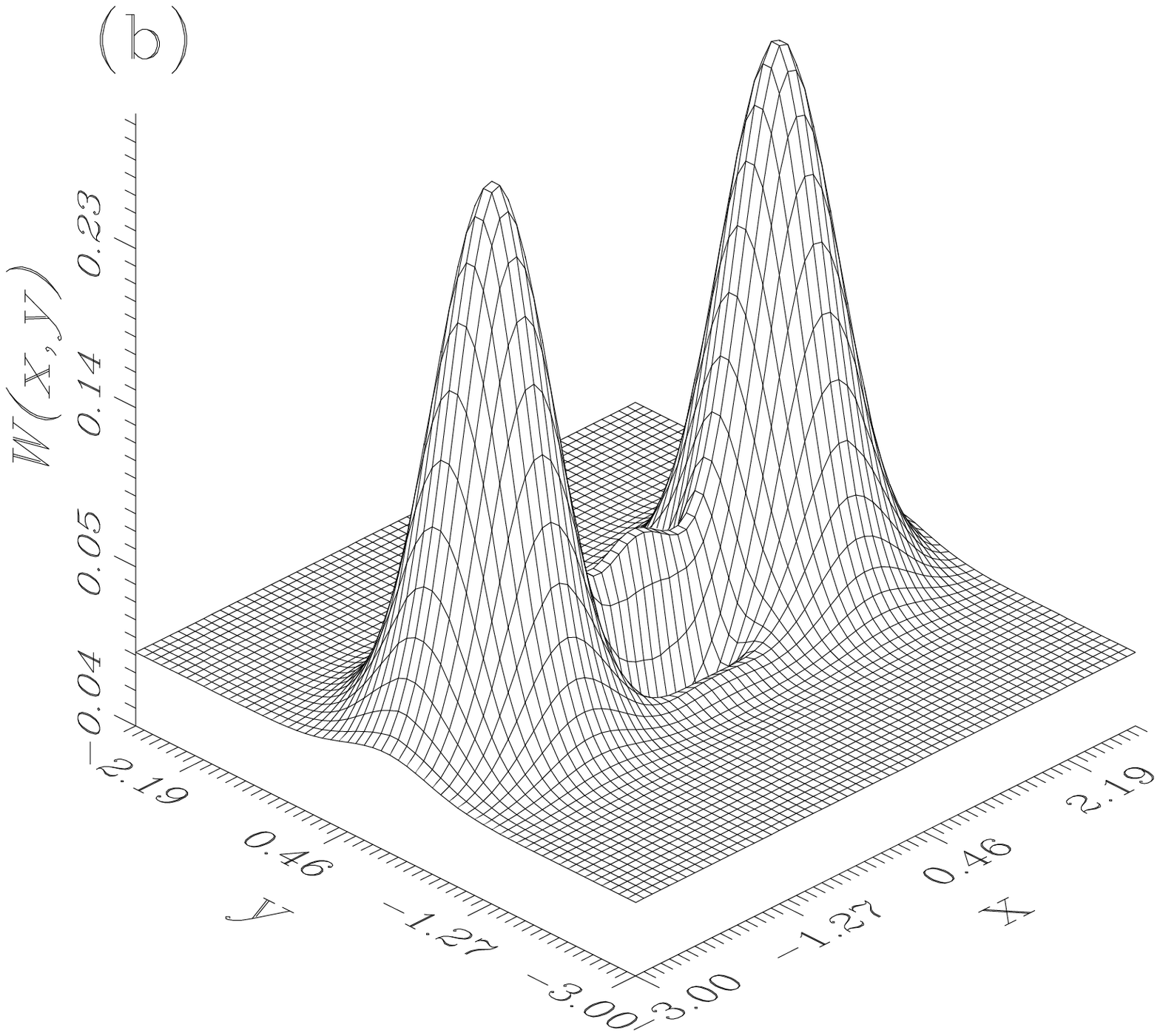}}
\caption{ The $W$ function of ECS case  for $(\alpha,
\tau)=(1,0.3)$ (a); $(\alpha, \tau)=(2,0.3)$ (b). }
\end{figure}
In general, the $W$ function of ECS, OCS and YSS (at $t=0$) are
consisting of two Gaussian bells corresponding to statistical
mixture of individual composite states and interference fringes in
between originating from the superposition between different
components of the states. Actually, these fringes represent the
signature of the nonclassical effects. For this reason several
articles have been devoted to deal with these fringes making them
less or more pronounced by allowing the cat states to evolve in
different quantum optical systems (e.g., see \cite{bivi}, and
references therein). For the system under consideration we can
easily conclude from
 (\ref{21b}) that as the interaction of the system with the
environment is going on, the two Gaussian peaks of the statistical-mixture
part move towards the origin and
eventually emerge into each other. This is quite obvious since the centers
of the peaks are exponentially decaying function of time. Furthermore, the amplitude of the
oscillatory term goes down by the factor $f(\alpha )$ similar to the $P(n)$.
Such behavior can be explained as
 the flux of coherent energy
transfers to the environment from the field  and noise transfers
to the field from the environment.
 More information about the system can be observed in Figs. 3a and b where
we  plot $W(\beta =x+iy)$ function for  microscopic ($\alpha=1 $)
and macroscopic ($\alpha=2 $) regimes, respectively.
In both cases the scaled decaying
parameter $\tau =0.3$. From Fig. 3b it is clear that the optical cavity
field tends to an approximate statistical mixture state, i.e.,
to a two-peak structure
 with
 negligible interference part. Actually the suppression of the nonclassical
interference pattern in the $W$ function does not mean that the
system reaches its equilibrium states \cite{dodonov}. Further for
large interaction times the cavity field collapses to  vacuum
state irrespective of the type of the initial cat state. This can
be checked from (\ref{21b}) as well as can be clearly seen in
Figs. 8 and 9 in \cite{buz1} (see curve 5 in these figures). This
means that the superpositions of macroscopical cat states can be
realized, but to have them surviving for some time the system must
be completely isolated. Even a very slight interaction with the
environment will very rapidly reduce the superpositions to the
corresponding statistical mixture states.

Now we turn our attention to the microscopic case (Fig. 3a). From this figure one
can observe that
the noise ellipse related to squeezed states is similar to
that of squeezed vacuum states. The origin of this behavior is in the
competition between the diagonal and off-diagonal elements of the system.
Actually, in the microscopic regime the contributions of the statistical mixture
components are located close to the origin of the phase space.
Furthermore, the comparison between the behavior of
 quadrature squeezing,
photon-number distribution  and $W$ function (i.e., the comparison
of
Fig. 1, Fig. 2 and Figs. 3a-b for the specified values of
the parameters) shows that the occurrence of the nonclassical effects and
decoherence phenomenon are qualitatively on the same level. More precisely, the more
the system decoheres, the more the nonclassical effects decrease. This
conclusion is completely different from that in \cite{buz1}. The reason of
this difference is that when the authors of \cite{buz1} compare the decay
of the interference part in phase space based on $W$ function (Fig. 6) with
the behavior of the quadrature squeezing (Figs. 8, 9) they chose for the
field amplitude $\alpha =2$, for which squeezing does not exist. Therefore
both developments cannot be compared and thus they arrived at misleading
conclusions. Furthermore, they explained their results by using series form
for the $W$ function (see (\ref{21a}); further
in Eq. (5.13) of \cite{buz1} there is a misprint in this expression, where
no square root should be in its denominator) and concluded that ''it is
clearly seen that the Wigner function always decays faster than the
second-order squeezing''. Actually, this discussion is not persuading
because the expansion contains the terms of both the mixture and the
interference components symmetrically.

\begin{figure}
\centering {\includegraphics[width=8cm]{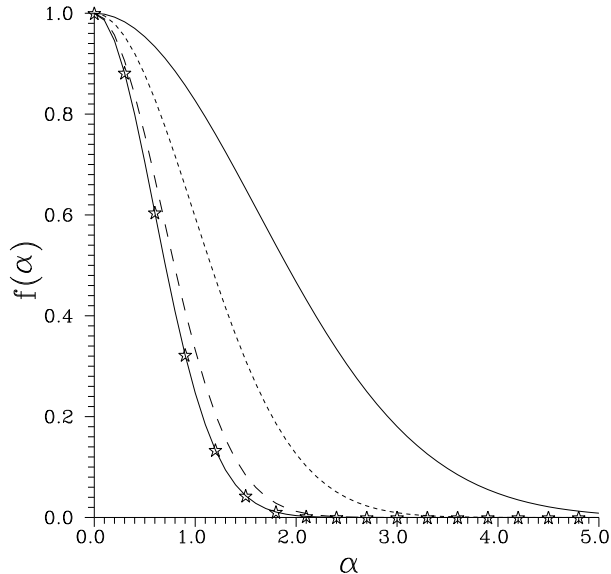}}
\caption{ The function $f(\alpha)$ aganist $\alpha$ for $%
\tau=0.1$ (solid curve), $0.3$ (short-dashed curve), $0.8$
(long-dashed curve) and $1.2$ (star-centered curve). }
\end{figure}

We conclude giving a quantitative analysis of the factor $%
f(\alpha )$ in Fig. 4. Such analysis can give insight into the occurrence
(or nonoccurrence) of the decoherence process regarding to the values of
 $\alpha $ and
the interaction time. As is clear from (\ref{4b}), $f(\alpha )$
exponentially decays whenever $\alpha $ increases provided that $\mu \neq 1$
(i.e., $\tau \neq 0$) and has its maximum value at $\mu =0$ (i.e., $\tau $
is very large). Actually, Fig. 4, even if it is relatively simple, it can give the
smallest values of $\alpha $ for which the system can be completely
decohered for  certain values of the interaction time. For instance, for
 $\tau =0.1,0.3,0.8$ the corresponding smallest values
are $\alpha =5,3,2$ for which the system is completely decohered. In this case
the density matrix describing the system has typically the form $\hat{\rho}%
_{{\it \tau }}(t)=\frac{1}{2}[|\alpha _{t}\rangle \langle \alpha
_{t}|+|-\alpha _{t}\rangle \langle -\alpha _{t}|]$ where $\alpha _{t}=\alpha
\sqrt{\mu }$ \cite{micro1}.  It is clear that these results agree with the fact
that the nonclassical effects occur in the microscopic regime. Finally, it
is worth mentioning that the decoherence in the present system can be
overcome by including amplifying media in the cavity \cite{aga}.

In conclusion, we have shown that the sensitivity of quadrature squeezing
and $W$ function to lossy mechanism is on the same level. This is not a
surprising result since the $W$ function is built on the complementarity of
the canonical operators \cite{ad1}. On the other hand, the $P(n)$ is more
sensitive to dissipation than the quadrature squeezing. Furthermore, the
decoherence process is more visible in the macroscopic regime. Thus in a more realistic situation the generation and detection of a
macroscopic superposition states is very difficult, due to the unavoidable
coupling with environment and the consequent dissipation \cite{zurek}.
Finally in the view of the quantities studied here, the nonclassical
superposition states cannot be saved from decoherence.

\acknowledgments
I  thank also Prof. J. Pe\v{r}ina and Dr. A.
Luk\v{s} from the Department of Optics, Palack\'y University,
Olomouc for their comments on the revised manuscript. Also I
acknowledge the support from the Project LN00A015 of Czech
Ministry of Education.

\end{document}